\documentclass[prc,aps,showpacs,floatfix,nofootinbib,twocolumn,letterpaper]{revtex4} 
\usepackage{graphicx} 
\usepackage{amssymb} 
\usepackage{amsmath} 
\usepackage{amsfonts}

\def\be{\begin{equation}} 
\def\ee{\end{equation}}

\def\eps{\varepsilon} 
\begin{document} 
\title{Monopole moments and nuclear compressibility
} 
\author{G.F.~Bertsch}
\affiliation{
Department of Physics and Institute of Nuclear Theory, 
Box 351560\\ University of Washington, Seattle, Washington 98195, USA} 
 
\begin{abstract} 
The nuclear compressibility has a role in nuclear physics in several
ways.  Its relationship to the giant monopole is well
known and has been subject of much theoretical work.  Less well
known is its affect on in nuclear structure, namely monopole transitions between 
low-lying states in the spectrum.  Here I revisit the relationship between
monopole and quadrupole moments in deformed nuclei.  It has often 
been assumed without any microscopic justification that the nucleus 
can be treated as an incompressible fluid.  An analytic formula
is derived here for the resulting relationship between monopole and 
quadrupole moments. The formula is shown to be well satisfied
within self-consistent mean-field theory calculated with several
energy functionals. 
The formula can be tested when both monopole and quadrupole matrix
elements of band-to-band transitions are known.  Data is available
for the decay of the first
excited $K^\pi = 0^+$ band to the ground-state band in $^{156}$Gd,
and the extracted matrix elements are found to be consistent with
the incompressible fluid picture.

\end{abstract} 
\maketitle 
 
\section{Introduction} 

P.F.~Bortignon had a great interest in the interplay between collective
and single-particle aspects of nuclear structure and their relation to
the nuclear Hamiltonian. Among the many topics was the giant monopole
vibration and its dependence on nuclear compressibility
\cite{ba88,va01,co01}.  In this note, I want to revisit the 
topic focusing on the low-lying monopole transitions in the
structure of deformed nuclei.  These transitions are prominent 
in the search for low-frequency $\beta$-vibrations, 
reviewed two decades ago by Garrett \cite{ga01}.

The existence of deformed nuclei has been known almost since the beginnings
of nuclear physics \cite{ca35}, but the theoretical understanding of 
nuclear shapes and sizes has greatly evolved over time.  It has been
commonly assumed that the nucleus is incompressible when undergoing shape 
changes \cite{re61,BMII}.  
However, rather simplified 
assumptions were made about the nucleon density distribution in those
early works.
A more general way to implement the incompressibility assumption 
is to treat the quadrupole shape changes
as taking place by an irrotational and divergence-free flow.  For axially
symmetric
deformations, a density distribution $\rho$ that starts out as 
spherically symmetric   $\rho(\vec r ) = \rho_0(|r|)$
is transformed to a deformed shape
$\rho(\vec r')$ by the scaling transformation
\be
\vec r' = (x',y',z') =  (x e^{\eps/2}, y e^{\eps/2}, z e^{-\eps}). 
\ee
Here $\eps$ is a deformation parameter.   The mean square radius
\be
 \langle r^2\rangle_\eps = \frac{1}{A}\int d^3 r\, r^2 \rho_0(\vec r')
\ee
of a deformed nucleus of mass number $A$ is easily expressed
in terms of the mean square radius $R^2_0$
for the spherical shape.  Namely,
\be
\label{r2}
\langle r^2 \rangle_\eps = \frac{1}{3}\left(e^{2\eps} + 2e^{-\eps}\right)R_0^2.
\ee
The corresponding moment of the mass quadrupole operator
\be
\hat Q^2 = z^2 -(x^2+y^2)/2 
\label{Q2def}
\ee 
can be written similarly as
\be
\label{Q2}
Q_2 \equiv \langle \hat Q_2 \rangle_\eps =  \frac{1}{3}\left(e^{2\eps} - e^{-\eps}\right)A R_0^2.
\ee
Combining Eqs. (\ref{r2}) and (\ref{Q2}), there is a 
parameter-free analytic relationship $F$
between $q = Q_2/A R_0^2$ and $\langle r^2\rangle_\eps/R_0^2$
which can be expressed 
\be
\langle r^2 \rangle_\eps = R_0^2 F(q).
\label{F0}
\ee
The formula for $F$ 
is derived in the Appendix as Eq. (\ref{Fz}).

\section{$\langle r^2 \rangle$ from self-consistent mean field theory} 

The next question is how well Eq. (\ref{F0}) is satisfied in
microscopic nuclear structure theory.  In particular,  self-consistent
mean-field (SCMF) theory is the method of choice for calculating
properties of heavy nuclei
including giant monopole vibrations.    For this study SCMF
is employed using
three different energy functionals: 
the Gogny D1S \cite{D1S},  the BCPM 
\cite{bcpm}, and the Skyrme SLy4 \cite{sly4}.  
The functionals are similar
in that they all require a density-dependent contact interaction to
achieve
nuclear saturation.  However, the details of the interaction are
rather different.  In the Gogny D1S and the
Skyrme SLy4 the form of the density dependence is a power law
$\rho^\alpha$ with
$\alpha=1/3$ for D1S and 1/6 for SLy4. In the BCPM
the density dependence is fitted to 
the equation of state of nuclear matter.
\begin{figure}[tb]
\begin{center}
\includegraphics[width=0.8\columnwidth]{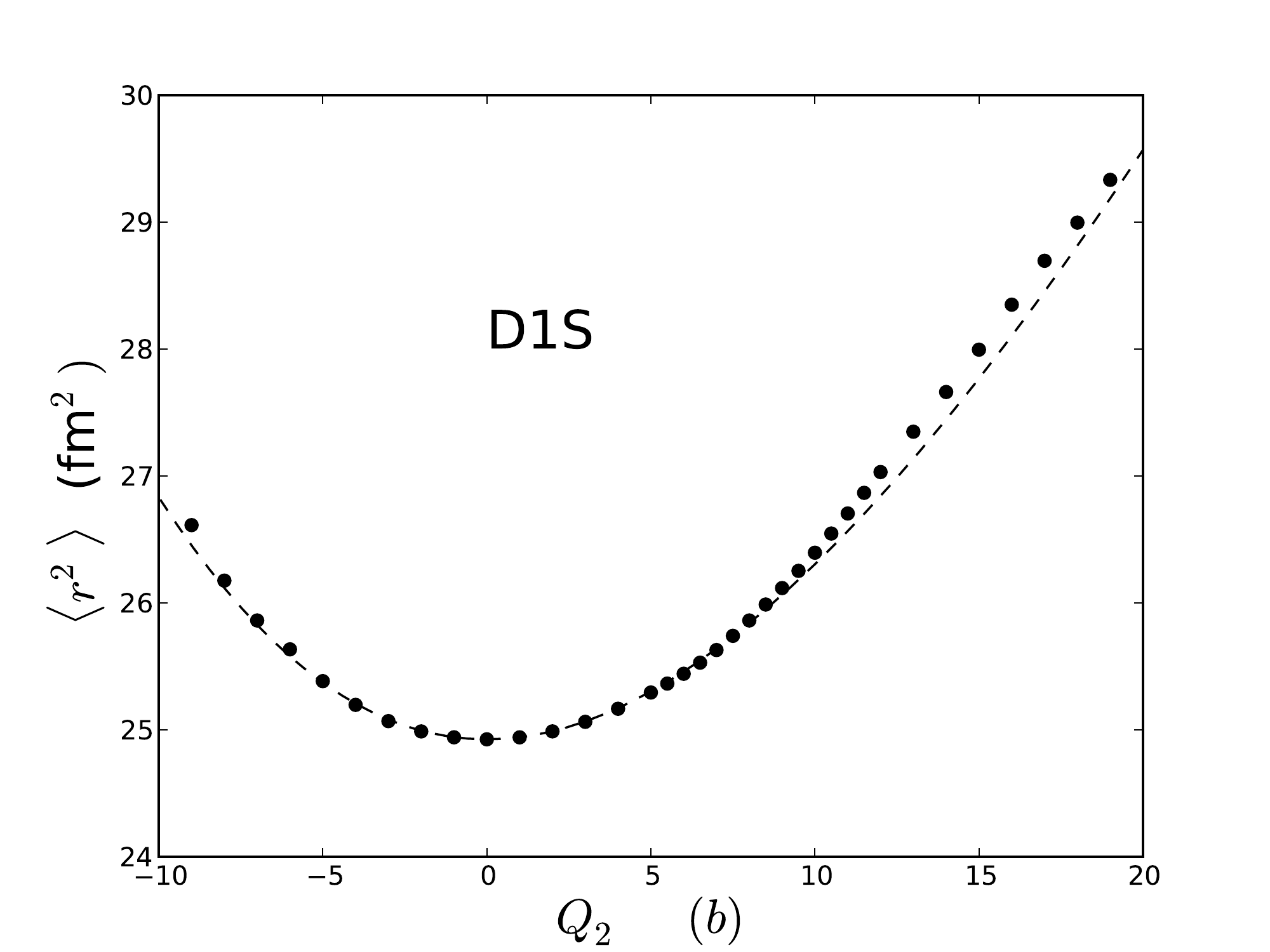}\\
\includegraphics[width=0.8\columnwidth]{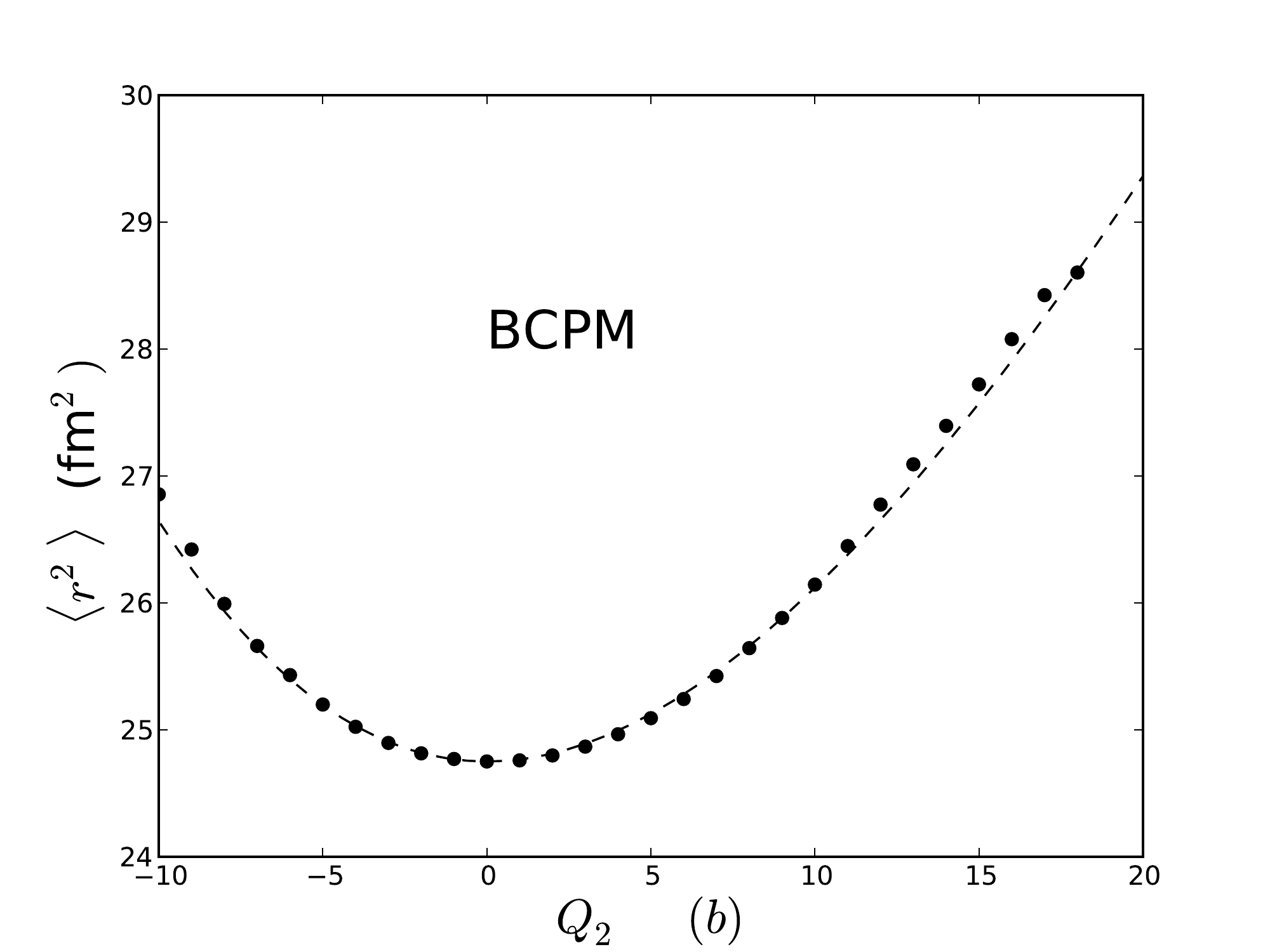}\\
\includegraphics[width=0.8\columnwidth]{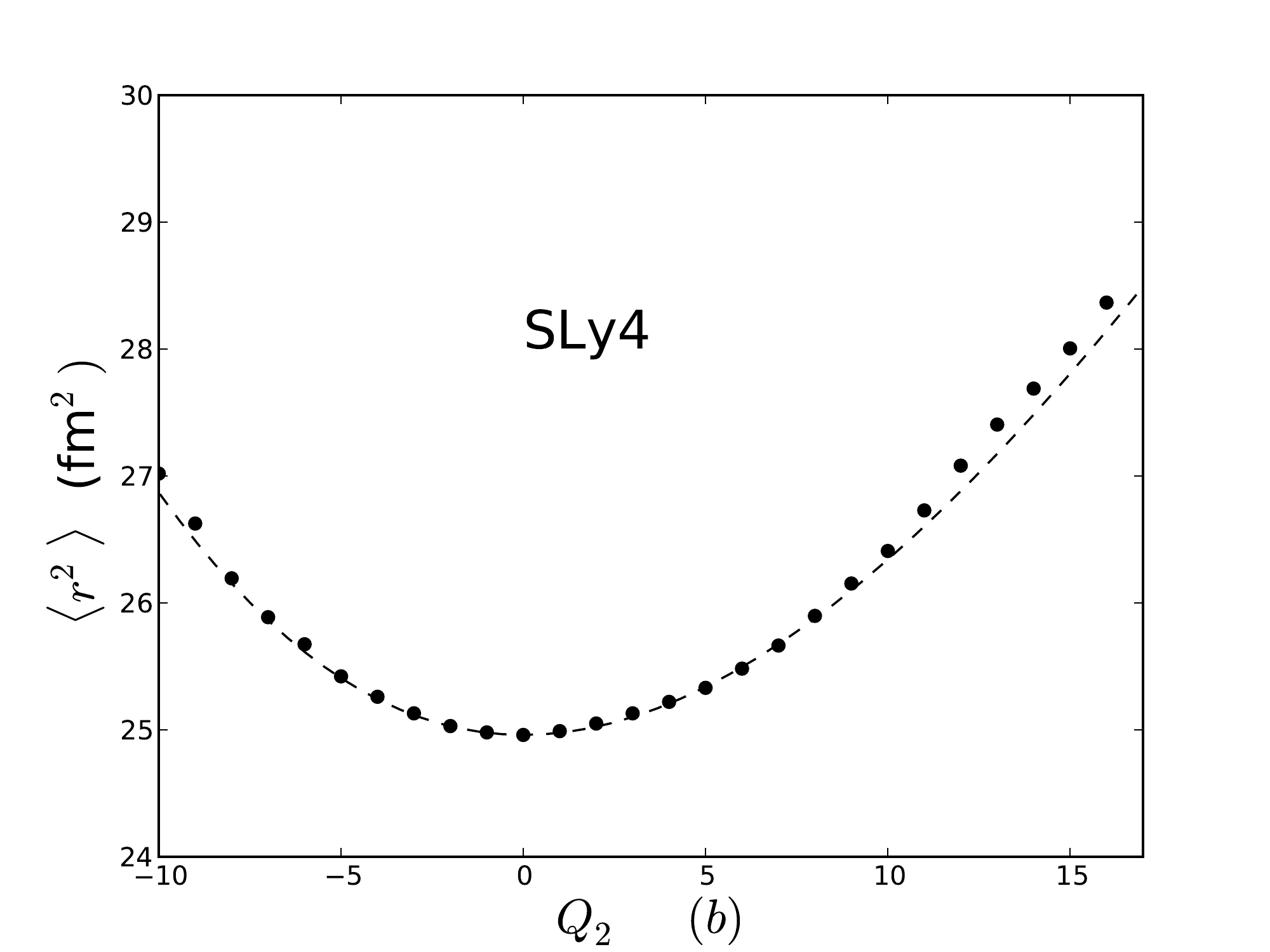}
\caption{Mean square radius in $^{156}$Gd as a function of mass
quadrupole moment.  The black circles show the radii calculated by
self-consistent mean field theory.
The dashed line shows the relationship
according to the incompressibility formula Eq. (\ref{Fz}).
\label{gd156} 
}
\end{center}
\end{figure}
Fluctuations in shape are accessed by the Generator Coordinate Method
(GCM), which involves minimizing the energy of the system while
constraining some shape variable(s).

The calculations were carried out for the nucleus
$^{156}$Gd, the subject of a recent experiment \cite{ap18}.  
The energy minimizations were performed in
the Hartree-Fock-Bogoliubov approximation
constraining the mass quadrupole moment.
The code {\tt HFBaxial}
\cite{robledo} was used for the D1S and the BCPM functionals,  while the SLy4 functional
was treated by the code {\tt ev8} \cite{bo05}.

Comparisons of calculated mean square charge radii with the incompressibility
formula are shown in Fig. 1.  
For all three functionals the mean square radii
are close to that predicted by Eq. (\ref{F0}).  This provides
a theoretical justification of the incompressibility 
assumption as applied to fluctuations in the quadrupole deformation.

\section{Transitions between bands}

Here I derive a formula for the ratio of monopole to quadrupole matrix
elements connecting two bands in a deformed nucleus.  It is 
conventional
to rescale the quadrupole field operator to a dimensionless operator
$\hat \beta$ as 
$
\hat \beta = \hat Q_2/k A R_0^2. 
$
where the proportionality constant is $k= 3/(20 \pi)^{1/2}$ and  
the spherical radius parameter is taken is $R_0= 1.2 A^{1/3}$ fm.
However, the scaling does not affect the physical relationships and so
I will use the $\hat Q_2$ directly in the present derivation.  The matrix elements to be calculated
are between the intrinsic wave functions of two bands $a$ and
$b$.  
The wave functions of both
bands are expanded in a basis of deformed Hartree-Fock (HF) configurations, 
and it is assumed 
that the $\hat Q_2$ operator is diagonal in that basis.   Then the
matrix element of $\hat Q_2$ can be expressed
\be
 \langle  a|\hat Q_{2} | b \rangle  =
\sum_k c_{ak}^* c_{bk} \langle k | \hat Q_2 | k \rangle
\ee
where $|k\rangle$ are the HF configurations and $c_{ak},c_{bk}$
are their amplitudes in the intrinsic states.  Under the
incompressibility assumption, the matrix elements of the monopole
operator are given by the formula
\be
 \langle  a| r^2 | b \rangle  = A  R_0^2 
\sum_k c_{ak}^* c_{bk} F\left(\langle k | \hat Q_2 | k \rangle/A R_0^2\right).
\label{r2fromQ2}
\ee
If the quadrupole moments of the HF components are all equal, the matrix
elements vanishes due to the orthogonality of the two bands.  Thus it
is only fluctuations in $\langle \hat Q_2 \rangle$ that permit a nonzero
monopole matrix element.  Recognizing that the fluctuation is small
compared to the average matrix element, the quadrupole operator
is written as the sum of the average and the fluctuation, 
\be
\hat Q_2 = \langle Q_2 \rangle + \delta \hat Q_2.
\ee

Inserting this into Eq.(\ref{r2fromQ2}), a first-order Taylor expansion
of $F$ yields
\be
\langle  a| r^2 | b \rangle  =  A  F'( \langle Q_2 \rangle /A R_0^2)
\langle a|\hat Q_2 | b \rangle
\label{ans}
\ee
where $F' = dF /dq$.
\section{Experiment}
To compare Eq. (\ref{ans}) with experiment, one has to extract the
band matrix elements from the 
measured spectroscopic transitions between levels.
For well-deformed systems, the relationship is
given in Eq. (4-68a) in Ref. \cite{BMII}.  If there is only one measured transition it is far 
from clear how well the band picture is satisfied.
However, recently a number of transitions were measured in  the nucleus
$^{156}$Gd \cite{ap18}.  There the band analysis was carried out for
seven quadrupole transitions between the first excited band $|0^+_2\rangle$
and the ground
state band $|g\rangle$.  The entire set of transitions was 
consistent with a single band-to-band 
matrix element of the charge quadrupole operator $\hat Q^{ch}_2 $, reported
as
\newcommand{\varA}[1]{{\operatorname{#1}}}
\be
\langle  g |\hat Q^{ch}_2 | 0^+_2 \rangle  = 30 \,\,\,\varA{e-fm}^2.
\ee
We may assume that the charge moments obey the same relation found for
the mass moments, replacing $A$ by $Z$ in Eq. (\ref{ans}).
We still need an estimate of intrinsic quadrupole moment of $|g\rangle$ to
apply Eq. (\ref{ans}).  In SCMF the moment is about 800 fm$^2$, giving
$q\approx 0.13$ and $F'(q) \approx 0.22$. 
Inserting this into Eq. (\ref{ans}), one obtains
\be
\langle  g | \hat r_{ch} ^2   | 0^+_2 \rangle = 6.6 \,\,\,\varA{e-fm}^2.
\ee  
For the experimental value, the tabulation \cite{ki05} gives a range
$\langle  g | \hat r_{ch} ^2   | 0^+_2 \rangle = 6 - 10$ fm$^2$.  
Thus
experiment is consistent with the incompressibility assumption, but the
uncertainties are too large to measure deviations from it.
It would require much more accuracy in the monopole measurement to make
a really strong test.  But it is certainly worth the effort
to learn more about nuclear compressibility.

\section{Acknowledgment}

The author thanks L.M. Robledo for providing the code HFBaxial used
for the calculations with the D1S and BCPM energy functionals.  The
author also thanks A. Aprahamian for discussions motivating the
present study.   

\vspace{0.5 cm}
\section{Appendix} 

\begin{figure}[tb]
\begin{center}
\includegraphics[width=0.8\columnwidth]{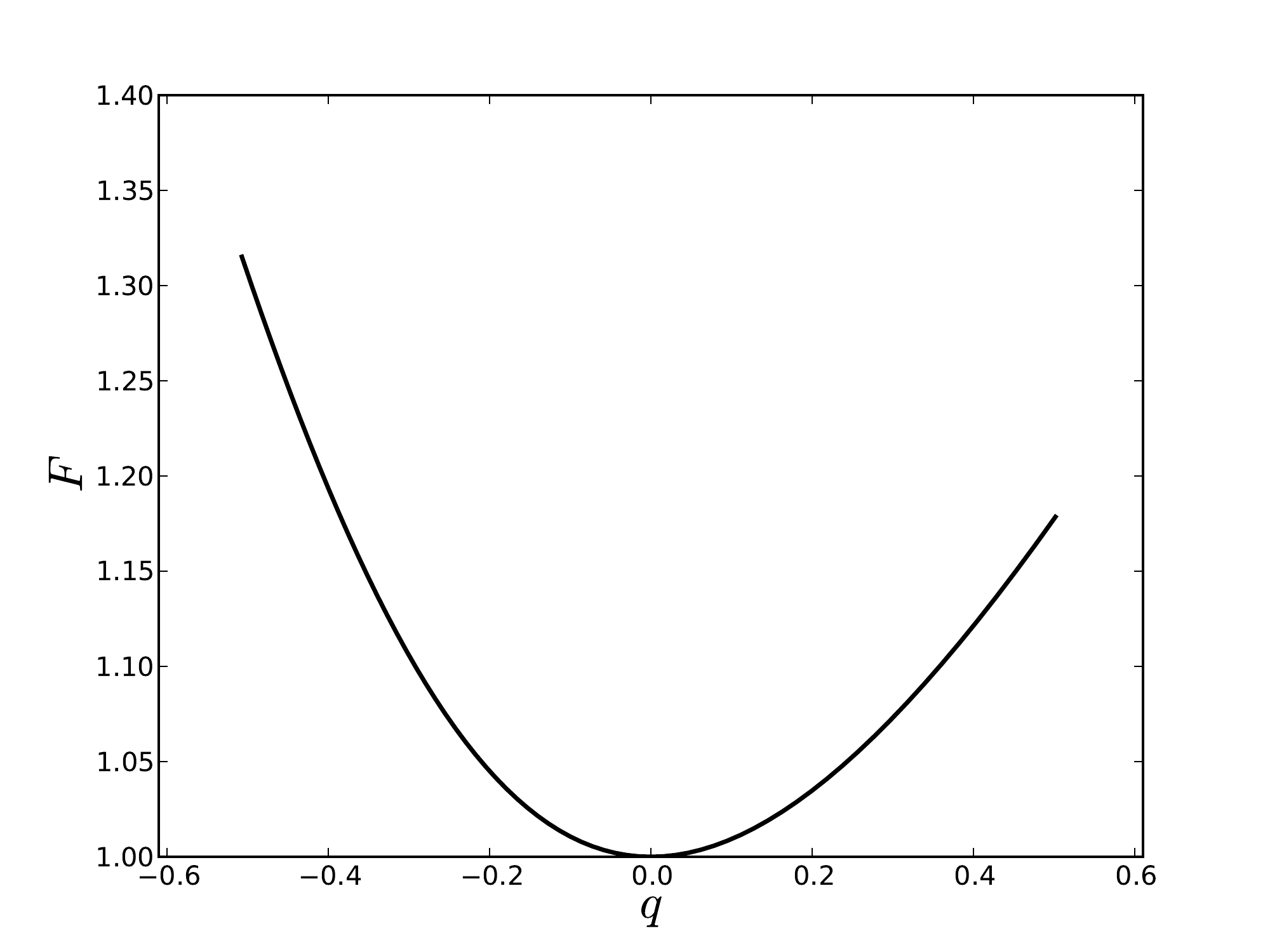}
\caption{The incompressibility function $F(q)$ in the range
$-1/2 < q < 1/2$.
\label{Ff} 
}
\end{center}
\end{figure}

The incompressibility  formula can be easily derived from Eq. 
(\ref{r2}) and
(\ref{Q2}).  First, change the deformation variable from $\eps$ to 
$w = e^{\eps}$.
Then Eq. (\ref{Q2}) is a cubic equation for $w$ in terms of $q = Q_2/A R^2_0$.  The
physical  solution is
\be
w = \frac{q}{D} + D
\ee
where 
\be
\label{A}
D = \left(\frac{1 +\sqrt{ 1 - 4q^3}}{2}\right)^{1/3}.
\ee
Substituting Eq. (\ref{A}) in Eq. (\ref{r2}) yields
\be
F(q) = \frac{1}{3} \left( \frac{q}{D}+ D\right)^2  + \frac{2}{3} 
\left( \frac{q}{D} +D \right)^{-1}.
\label{Fz}
\ee
The function $F(q)$ is shown in Fig. \ref{Ff}.

\end{document}